\documentclass[11pt]{article}
\usepackage{graphicx}

\begin{document}

\title{Finite Temperature Phase Diagram in Rotating Bosonic Optical Lattice}

\author{Beibing Huang and Shaolong Wan\thanks{Corresponding author.
Electronic address: slwan@ustc.edu.cn} \\
Institute for Theoretical Physics and Department of Modern Physics \\
University of Science and Technology of China, Hefei, 230026, {\bf
P. R. China}} \maketitle
\begin{center}
\begin{minipage}{120mm}
\vskip 0.8in
\begin{center}{\bf Abstract} \end{center}

{Finite temperature phase boundary between superfluid phase and
normal state is analytically derived by studying the stability of
normal state in rotating bosonic optical lattice. We also
analytically prove that the oscillation behavior of critical hopping
matrix directly follows the upper boundary of Hofstadter butterfly
as the function of effective magnetic field.}

\end{minipage}
\end{center}

\vskip 1cm

\textbf{PACS} number(s): 03.75.Lm, 05.30.Jp, 73.43.Nq
\\

\section{Introduction}

Bose-Hubbard model of interacting bosons on a lattice has been used
to describe superfluid-Mott insulator (MI) phase transition in a
variety of systems at zero temperature, e.g., Josephson arrays and
granular superconductors \cite{fisher}. The recent suggestion to
experimentally observe this transition in a system of cold bosonic
atoms in an optical lattice \cite{jaksch} and its successful
experimental demonstration \cite{markus} have aroused much
theoretical \cite{bpdas, santos, polak} and experimental
\cite{mckay, bloch} interest in this model, especially rotating
optical lattice has also become brand-new topics in bosonic system.
Most of work about rotating optical lattice focused on the
superfluid phase and studied the pinning effect of the vortex
lattice due to optical periodic potential \cite{kenichi, rajiv,
h.pu, j.w.}. The similar question has been investigated in the
type-$\Pi$ superconductor \cite{martin, david, maniv, wvpogosov}.

However the question of superfluid-MI phase transition at zero
temperature still exists in rotating optical lattice and the phase
diagram at zero temperature has been achieved by strong coupling
expansion \cite{niemeyer} and Gutzwiller approach \cite{oktel1,
oktel2}. Strong coupling expansion obtained the phase boundary
between the superfluid phase and MI by perturbatively computing the
energy difference between MI and single-charge excitation states on
top of the MI. This method is very accurate and applicable to the
random dimension but is not suitable for system in deep superfluid
phase where the perturbation is not valid any longer. Gutzwiller
approach is at the self-consistent mean-field level and based on an
ansatz that many-body ground state factorizes into product of single
lattice site wave function. So under this approximation the system
become diagonal with respect to the lattice site and we can use an
effective single-site Hamiltonian. The disadvantage of Gutzwiller
approach is that it fails to describe the correct short-range
correlation between different lattice sites and so is an
uncontrolled approximation. However, Gutzwiller approach can predict
a qualitatively similar phase diagram with strong coupling expansion
\cite{oktel1}.

In this paper, we mainly extend the phase diagram of a rotating
two-dimensional bosonic optical lattice to finite temperature
utilizing the Gutzwiller mean-field theory. At finite temperature,
MI is replaced by normal state which possesses finite
compressibility. Here we do not include the crossover from MI to
normal state at finite temperature since there is not an
conventional definition for this crossover as far as what we know is
concerned. So at finite temperature the phase transition happens
between superfluid phase and normal state instead of MI. In section
2, by making an analogy between rotating optical lattice and
electrons constrained by periodic potential and external magnetic
field, we qualitatively derive the Bose-Hubbard Hamiltonian in
rotating optical lattice. In section 3, we follow the method used in
\cite{vezzani} to analytically locate the phase boundary of
superfluid phase and normal state by discussing the stability of
fixed point corresponding to the normal state and in section 4 a
brief conclusion is given.

\section{Bose-Hubbard Hamiltonian in Rotating Optical Lattice}

We consider a two-dimensional bosonic system in $XY$ plane
restricted by a square optical lattice and an harmonic trapping
potential which have a common rotating velocity $\Omega$ along $Z$
axis. In the laboratory frame, the potential is generally
time-dependent. It is therefore convenient to transform to the frame
rotating with the potential, since in that frame the potential is
constant in time, and thus the standard methods for finding the
equilibrium may be employed. In the frame of rotating potential, the
second quantized Hamiltonian for a particle of mass $m$ in an
harmonic trap of natural frequency $\omega$ can be written into
$H=H_0+H_I$ with
\begin{eqnarray}
H_0 &=& \int d\,\vec{R}\, \Psi^{\dag}(\vec{R})\left(-\frac{\hbar^2
\nabla^2}{2m}+ V_L(\vec{R})+\frac{1}{2}m \omega^2 R^2- \Omega L_Z -
\mu
\right)\Psi(\vec{R}) \label{1} \\
H_I &=& \frac{g}{2}\int d\,\vec{R}\,\Psi^{\dag}(\vec{R})
\Psi^{\dag}(\vec{R}) \Psi(\vec{R}) \Psi(\vec{R})  \label{2}
\end{eqnarray}
where $L_Z=-i\hbar(X\partial_Y-Y\partial_X)$ is angular momentum
along $Z$ axis and $g(>0)$ is the strength of contact interaction
potential between two particles. $V_L(\vec{R})$ is periodic optical
potential and $\Psi(\vec{R})$ is field operator for boson particles.
The term involving the chemical potential $\mu$ is added because it
is very convenient to be in the grand canonical ensemble. $H_0$ can
be rearranged into
\begin{eqnarray}
H_0=\int d\,\vec{R}\, \Psi^{\dag}(\vec{R})
\left[\frac{(-i\hbar\nabla-e/c\vec{A}(\vec{R}))^2}{2m}+V_L(\vec{R})+
\frac{1}{2}m (\omega^2-\Omega^2) R^2- \mu \right]\Psi(\vec{R})
\label{3}
\end{eqnarray}
where $\vec{A}(\vec{R})=mc/e\vec{\Omega}\times \vec{R}$ is an
effective vector potential with $c$ and $e$ representing light speed
and charge quanta. This form suggests that the effects of rotation
are partitioned into two different parts. The term $1/2m
(\omega^2-\Omega^2) R^2$ implies that the centrifugal potential
weakens the role of trapping potential (in order to stabilize the
system, $\Omega \leq \omega$). The other part of rotation is
included in the first term whose role is producing an effective
magnetic field $\vec{B} =\nabla \times \vec{A}(\vec{R})=
2mc/e\vec{\Omega}$ and provides a structure of Landau level for
atoms. Hence at this point we can draw a conclusion that the motion
of atoms in a rotating optical lattice under the assumption
$\omega=\Omega$ is completely the same as that of electrons
constrained by periodic potential and external magnetic field.
Therefore the method utilized to deal with electrons can be
applicable to the our problem. For simplicity, below we assume
$\omega=\Omega$ so that the centrifugal force accurately compensates
the harmonic trapping potential.

For systems without rotation and trap potential ($\Omega=\omega=0$)
we usually assume that the atoms are cooled within the lowest Bloch
band and the field operator can be expanded into $\Psi(\vec{R}) =
\sum_{i} b_{i} w(\vec{R} - \vec{R}_i )$ in terms of the lowest
Wannier function $w(\vec{R}- \vec{R}_i)$ with $b_{i}$ being bosonic
annihilation operator at the lattice site $\vec{R}_i$. Hence the
system is sufficiently described by a single-band Bose-Hubbard
Hamiltonian \cite{jaksch}
\begin{eqnarray}
H &=& -t\sum_{<ij>}b_i^{\dag}b_j + \frac{U}{2} \sum_{i} b_{i}^{\dag}
b_{i}^{\dag} b_{i} b_{i} - \mu \sum_{i} b_{i}^{\dag} b_{i} \label{4}
\end{eqnarray}
where $t=-\int d \vec{R}\,w^{\ast}(\vec{R} -
\vec{R}_i)\left[-\frac{\hbar^2 \nabla^2}{2m}+
V_L(\vec{R})\right]w(\vec{R} - \vec{R}_j)$ is hopping matrix
restricted to nearest neighbors and $U=g\int d \vec{R}\,|w(\vec{R} -
\vec{R}_i)|^4$ is on-site interaction strength. When external
effective magnetic field appears, many-band effects must be
considered which leads to that single-band Bose-Hubbard Hamiltonian
is not valid any more. Fortunately from the study of lattice
electron in external magnetic field \cite{fabian} we know that there
exists an effective single-band Hamiltonian which can be obtained
from (\ref{4}) by only making a substitution for hopping matrix
\begin{eqnarray}
H=-t\sum_{<ij>}\exp{\left[\frac{ie}{\hbar c}
\int_{\vec{R}_j}^{\vec{R}_i} \vec{A}(\vec{R})\cdot d
\vec{R}\right]}b_i^{\dag}b_j + \frac{U}{2} \sum_{i} b_{i}^{\dag}
b_{i}^{\dag} b_{i} b_{i} - \mu \sum_{i} b_{i}^{\dag} b_{i} \label{5}
\end{eqnarray}
The added phase factor is called Peierls phase factor. It is well
known that superfluid-MI phase transition comes from the competition
between hopping matrix and on-site interaction strength
\cite{markus}. But when the magnetic field is introduced the on-site
interaction strength is unchanged while the hopping matrix is
modified, so we naturally expect a modified phase boundary between
superfluid and MI.

The vector potential $\vec{A}(\vec{R})$ in the symmetric gauge
concerns the $X$ and $Y$ components of coordinate at the same time
and makes below calculation be more complex. We hope only $X$ or $Y$
component is concerned, which is realized by making a canonical
transformation to field operator
\begin{eqnarray}
b_i \longrightarrow b_i \exp{\left[\frac{ie}{\hbar c}
\int_{\vec{R}_i}^{\vec{R}_l} mc/e\vec{\Omega}\times
\vec{\underline{R}}\cdot d \vec{R}\right]} \label{6}
\end{eqnarray}
with $\vec{\underline{R}}=(X, -Y)$ and $\vec{R}_l$ being a arbitrary
reference point. Under this transformation
\begin{eqnarray}
H=-t\sum_{<ij>}\exp{\left[i 2\pi \phi \int_{\vec{r}_j}^{\vec{r}_i} x
d y\right]}b_i^{\dag}b_j + \frac{U}{2} \sum_{i} b_{i}^{\dag}
b_{i}^{\dag} b_{i} b_{i} - \mu \sum_{i} b_{i}^{\dag} b_{i} \label{7}
\end{eqnarray}
where all the coordinates are scaled by the lattice constant $a$
hence are dimensionless. $\phi=Ba^2/(hc/e)$ represents the number of
magnetic flux quanta penetrating the unit cell. In fact it is easy
to find that the above Hamiltonian corresponds to that in Landau
gauge $\vec{A}(\vec{R})=BX\hat{Y}$ with $\hat{Y}$ denoting the unit
vector along the $Y$ axis. In the next section, we will regard the
Hamiltonian (\ref{7}) as our starting point and study its phase
diagram at finite temperature at mean-field level.

\section{Phase Diagram in Gutzwiller Mean-Field Approach}

The Gutzwiller approach is a self-consistent mean-field method and
equivalent to the decoupling approximation \cite{sheshadri, van} to
the hopping term
\begin{eqnarray}
b_{i}^{\dag} b_{j} &=& <b_{i}^{\dag}> b_{j} + b_{i}^{\dag} <b_{j}> -
<b_{i}^{\dag}> <b_{j}>
\nonumber \\
&=& \alpha_{i} b_{j} + b_i^{\dag} \alpha_{j} - \alpha_{i} \alpha_{j}
\label{8}
\end{eqnarray}
where $\alpha_i=\alpha_i^{\ast}$ is superfluid order parameter which
distinguishes the superfluid phase from normal state. If magnetic
field vanishes, the whole system is uniform and order parameter is
also site-independent. But we can not suppose this when the magnetic
field appears. After this decoupling, the system is describable in
terms of a single-site Hamiltonian
\begin{eqnarray}
H_{nm}&=&-t\left[b_{nm}^{\dag} \left(\alpha_{(n+1)m}+
\alpha_{(n-1)m}+ e^{-i2\pi n \phi} \alpha_{n(m+1)}+ e^{i2\pi n \phi}
\alpha_{n(m-1)}\right)
+ H. C.\right] + \nonumber \\
&&\frac{U}{2} b_{nm}^{\dag} b_{nm}^{\dag} b_{nm} b_{nm} - \mu
b_{nm}^{\dag} b_{nm} \label{9}
\end{eqnarray}
where we label site of the lattice $i$ by two ordered integers
$i=(n,m)$, the first integer along the $X$ axis and the second one
along the $Y$ axis. In the Landau gauge, hopping along the $Y$ axis
achieves the Peierls phase factor and that along $X$ axis is
invariant. The above Hamiltonian has two striking characteristics
\cite{oktel1}. On the one hand, it is independent of $Y$ component,
so the translational symmetry along $Y$ axis is conservative and we
can suppose order parameter $\alpha_{nm}=\alpha_n$ in correspondence
with the case without magnetic field. On the other hand although it
depends on $X$ component, for rational $\phi=p/q$ ($p, q$ have no
common factor), q-site translational symmetry along $X$ axis is
recovered $\alpha_{n}=\alpha_{n+q}$. So the Hamiltonian is further
reduced into
\begin{eqnarray}
H_{n}&=&-t\left[b_{n}^{\dag} \left(\alpha_{n+1}+ \alpha_{n-1}+
2\alpha_{n}\cos{2\pi n \phi}\right) + H. C.\right] + \frac{U}{2}
b_{n}^{\dag} b_{n}^{\dag} b_{n} b_{n} - \mu b_{n}^{\dag} b_{n}
\label{10}
\end{eqnarray}
with $n$ being integer from $1$ to $q$. In addition, the Hamiltonian
is periodic as the function of magnetic field
$H_{n}(\phi)=H_n(\phi+K)$ with $K$ being a random integer so that we
only need consider $\phi \in[0,1)$. Note also that in (\ref{9}) and
(\ref{10}) we have neglected a constant term which does not
influence our result.

The self-consistency of Gutzwiller method must be carried out by the
condition
\begin{eqnarray}
\alpha_{n}=\frac{1}{Z_n}Tr \left(b_{n}e^{-\beta H_{n}}\right)
=\frac{1}{Z_n}Tr\left( b_{n}^{\dag}e^{-\beta H_{n}}\right)
\label{11}
\end{eqnarray}
with $\beta=1/(K_BT)$ and partition function $Z_n=Tr \exp{(-\beta
H_{n})}$. Introducing the same notation in \cite{vezzani}
$\gamma_n=t\alpha_n$, self-consistent condition can be rewritten
into
\begin{eqnarray}
\gamma_n=\frac{t}{6\beta}\left(\frac{\partial}{\partial
\gamma_{n+1}} + \frac{\partial}{\partial \gamma_{n-1}} +
\frac{1}{2\cos{2\pi n \phi}}\frac{\partial}{\partial
\gamma_{n}}\right)\ln{Z_n} \label{12}
\end{eqnarray}
Under Gutzwiller approximation the eigenstates of $H_n$ can be
expanded in terms of Fock state, then we diagonalize this
Hamiltonian matrix under Fock basis truncated until a given number
of particle $N$ to obtain the eigenvalues $E_l$ ($l=0, 1,
\cdot\cdot\cdot, N$) which are function of $\gamma_{n-1},
\gamma_{n}$ and $\gamma_{n+1}$. So we proceed using the eigenvalues
\begin{eqnarray}
\gamma_n=-\frac{t}{6}\sum_{l=0}^{N}e^{-\beta E_l}\left(
\frac{\partial E_l}{\partial \gamma_{n-1}} +\frac{\partial
E_l}{\partial \gamma_{n+1}}+ \frac{1}{2\cos{2\pi n \phi}}
\frac{\partial E_l}{\partial
\gamma_{n}}\right)/\sum_{l=0}^{N}e^{-\beta E_l} \label{13}
\end{eqnarray}
The derivative of eigenvalue is computable from the characteristic
polynomial $P(\lambda, \gamma_{n-1}, \gamma_{n}, \gamma_{n+1})$ of
Hamiltonian matrix \cite{meyer}
\begin{eqnarray}
\frac{\partial E_l}{\partial \gamma_i}&=&-\frac{\partial P(E_l,
\gamma_{n-1}, \gamma_{n}, \gamma_{n+1})}{\partial \gamma_i}/
\frac{\partial P(E_l, \gamma_{n-1}, \gamma_{n},
\gamma_{n+1})}{\partial E_l}\nonumber \\
\frac{\partial P(E_l, \gamma_{n-1}, \gamma_{n},
\gamma_{n+1})}{\partial E_l}&=&\sum_{k=0}^{N} P^{(k)}(E_l,
\gamma_{n-1},
\gamma_{n}, \gamma_{n+1}) \nonumber \\
\frac{\partial P(E_l, \gamma_{n-1}, \gamma_{n},
\gamma_{n+1})}{\partial \gamma_{n-1}}&=&2(\gamma_{n-1}+\gamma_{n+1}
+2\gamma_n \cos{2 \pi n \phi})\cdot \nonumber \\
&&\left[\sum_{k=1}^{N} k P^{(k, \,k-1)}(E_l, \gamma_{n-1},
\gamma_{n}, \gamma_{n+1})+ Q(\gamma_{n-1}, \gamma_{n},
\gamma_{n+1})\right] \nonumber \\
\frac{\partial P(E_l, \gamma_{n-1}, \gamma_{n},
\gamma_{n+1})}{\partial \gamma_{n+1}}&=& \frac{\partial P(E_l,
\gamma_{n-1}, \gamma_{n}, \gamma_{n+1})}{\partial \gamma_{n-1}}
\nonumber \\
\frac{\partial P(E_l, \gamma_{n-1}, \gamma_{n},
\gamma_{n+1})}{\partial \gamma_{n}}&=& 2 \cos{2 \pi n \phi}
\frac{\partial P(E_l, \gamma_{n-1}, \gamma_{n},
\gamma_{n+1})}{\partial \gamma_{n-1}} \label{14}
\end{eqnarray}
where the notation $P^{\{k\}}(E_l, \gamma_{n-1}, \gamma_{n},
\gamma_{n+1})$ denotes the characteristic polynomial of the matrix
obtained by discarding from Hamiltonian matrix the rows and columns
labeled by the set of indices $\{k\}$ and the polynomial
$Q(\gamma_{n-1}, \gamma_{n}, \gamma_{n+1})$ satisfies $Q(0, 0, 0) =
0$. Substituting all above relations into (\ref{13})
\begin{eqnarray}
\gamma_n&=&t(\gamma_{n-1}+\gamma_{n+1} +2\gamma_n \cos{2 \pi n
\phi}) \frac{\sum_{l=0}^{N} e^{-\beta E_l}W(E_l, \gamma_{n-1},
\gamma_{n},
\gamma_{n+1})} {\sum_{l=0}^{N} e^{-\beta E_l}} \nonumber \\
W(E_l, \gamma_{n-1}, \gamma_{n},
\gamma_{n+1})&=&\frac{\sum_{k=1}^{N} k P^{(k,\, k-1)}(E_l,
\gamma_{n-1}, \gamma_{n}, \gamma_{n+1})+ Q(\gamma_{n-1}, \gamma_{n},
\gamma_{n+1})}{\sum_{k=0}^{N} P^{(k)}(E_l, \gamma_{n-1}, \gamma_{n},
\gamma_{n+1})} \label{15}
\end{eqnarray}
Until now, we have obtained the equation set of self-consistently
deciding all order parameters. If order parameters have nonzero
solution the system is in superfluid phase. If order parameters have
zero solution, the system is in normal state. So we can determine
the phase boundary between superfluid phase and normal state by
studying the stability of the fixed point corresponding to the
normal state ($\gamma_n=0$ for all $n$) \cite{vezzani}. According to
a standard theory, the stability of such a fixed point can be
discussed based on the spectrum of the matrix linearizing the map
defined by equation set (\ref{15}) in the vicinity of normal state.
Linearizing (\ref{15}) around the fixed point of normal state, we
obtain
\begin{eqnarray}
\gamma_n&=&t\Theta(U, \mu, \beta)(\gamma_{n-1}+2\gamma_n \cos{2\pi n
\phi}+\gamma_{n+1}) \label{16} \\
\Theta(U, \mu,
\beta)&=&\frac{\sum_{l=0}^{N}e^{-\beta\left(U/2l(l-1)-\mu
l\right)}W(U/2l(l-1)-\mu l, 0, 0, 0)}
{\sum_{l=0}^{N}e^{-\beta\left(U/2l(l-1)-\mu l\right)}} \label{17}
\end{eqnarray}
Writing above equations in the form of matrix if we denote
$\Upsilon=(\gamma_1, \gamma_2, \cdot \cdot \cdot \gamma_q)^{t}$
\begin{eqnarray}
\Upsilon&=&t\Theta(U, \mu, \beta)M(\phi)\Upsilon \nonumber \\
M(\phi)&=&\left(
\begin{array}{cccccc}
2\cos{2\pi \phi} & 1  & 0 & \cdot \cdot \cdot & 0 &1\\
\smallskip
 1 & 2\cos{4\pi \phi} & 1 & \cdot \cdot \cdot & 0 &0\\
\smallskip
\cdot\cdot\cdot&\cdot\cdot\cdot&\cdot\cdot\cdot&\cdot\cdot\cdot&\cdot\cdot\cdot&\cdot\cdot\cdot\\
\smallskip
 1    &  0    &  0  & \cdot\cdot\cdot & 1 & 2\cos{2\pi q \phi}
\end{array}
\right) \label{18}
\end{eqnarray}
Therefore, the fixed point of the normal state is stable only if the
maximal eigenvalue $\epsilon_{max}(\phi)$ of $M(\phi)$ satisfies
$t\Theta(U, \mu, \beta)\epsilon_{max}(\phi)<1$, which signifies that
the phase transition between superfluid phase and normal state is
\begin{eqnarray}
t\Theta(U, \mu, \beta)\epsilon_{max}(\phi)=1 \label{19}
\end{eqnarray}
This is our main result. At $T=0K$ only the eigenstate having the
lowest energy contributes to the partition function. Easily proven
when $\mu \in [l-1, l]$, the phase boundary at $T=0K$ is specified
by
\begin{eqnarray}
t=\frac{1}{\epsilon_{max}(\phi)} \frac{[\mu-U(l-1)][U l-\mu]}{U+\mu}
\label{20}
\end{eqnarray}
which is the same as the result in \cite{oktel2, goldbaum}.

Below we connect $\epsilon_{max}(\phi)$ with the famous Hofstadter
butterfly. We find that eigenvalues of the matrix $M(\phi)$ actually
correspond to a part of energy spectrum of Hofstadter butterfly
\cite{hofstadter}. According to the proof in \cite{oktel2} that the
maximal eigenvalue of $M(\phi)$ is equal to the maximal eigenvalue
of Hofstadter butterfly, hence from (\ref{19}) we find for fixed $U,
\mu, \beta$ the critical hopping strength is inversely proportional
to the maximal eigenvalue of Hofstadter butterfly. According to the
fact that the maximal eigenvalue of Hofstadter butterfly shows an
oscillatory behavior as the function of magnetic field
\cite{hofstadter}, the critical hopping strength also exhibits the
oscillation following the maximal eigenvalue, i.e. upper boundary of
Hofstadter butterfly.

\begin{figure}[htbp]
\includegraphics[width=9.0cm, height=6.0cm]{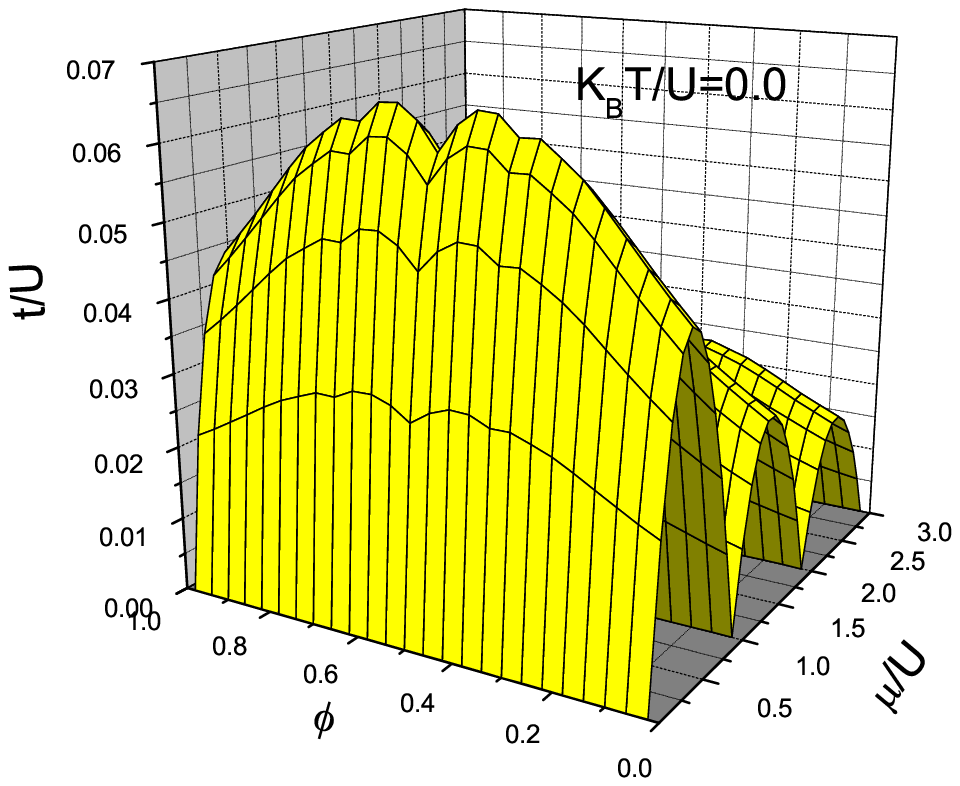}
\includegraphics[width=9.0cm, height=6.0cm]{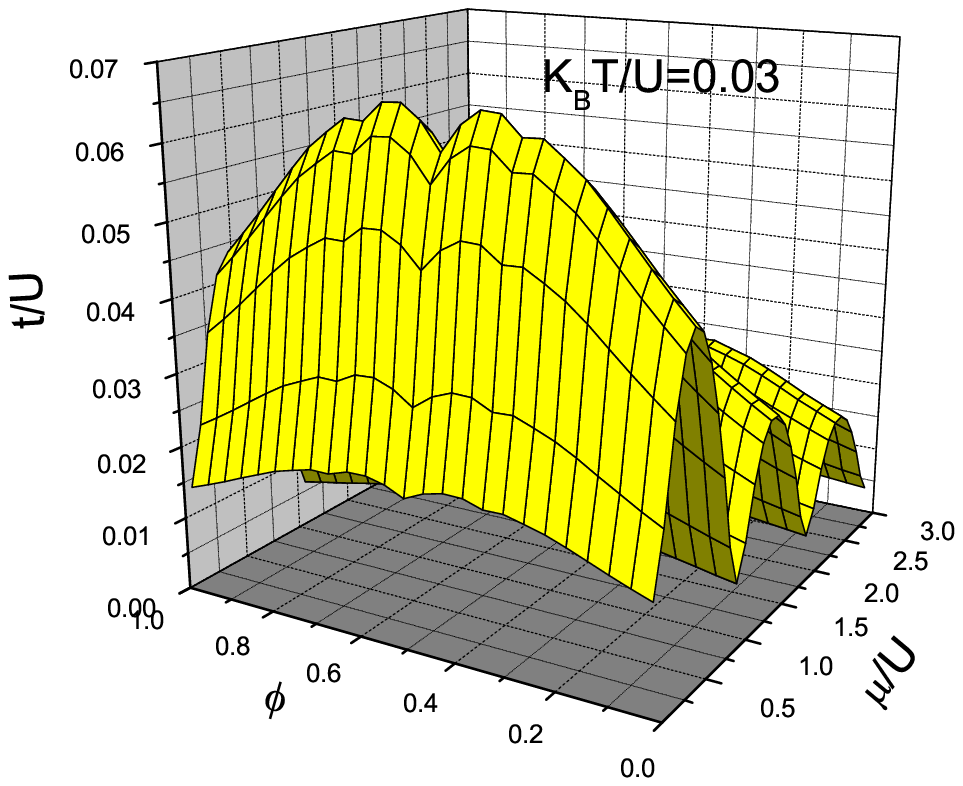}
\includegraphics[width=9.0cm, height=6.0cm]{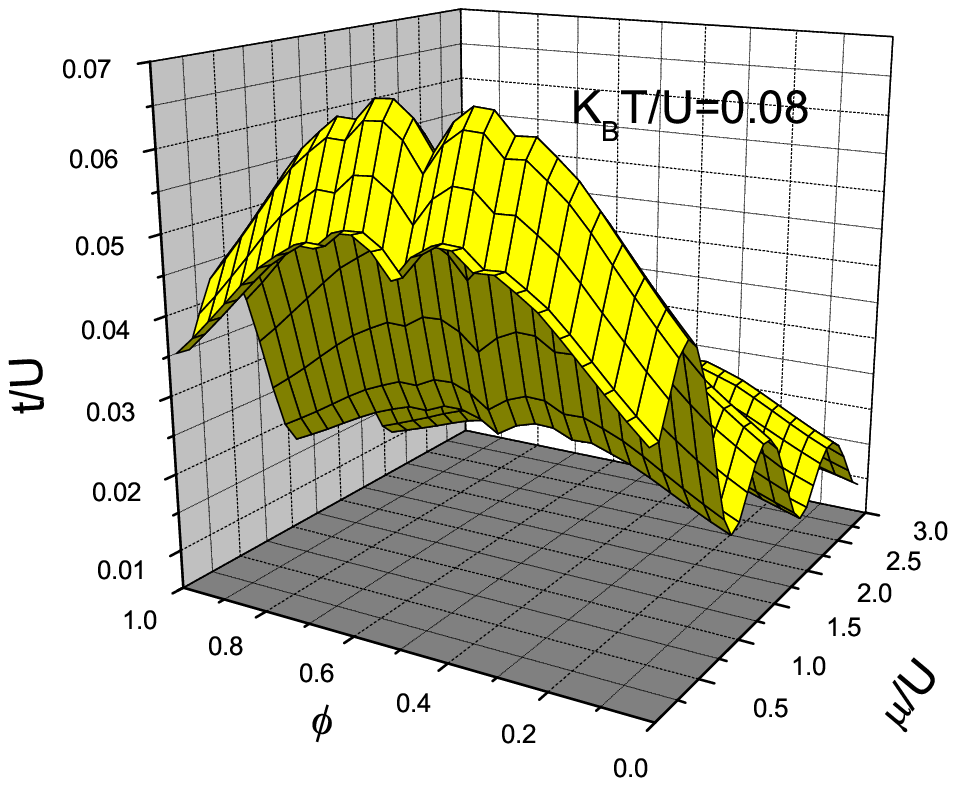}
\caption{The three-dimensional phase diagram for different
temperature $K_BT/U=0.0, 0.03, 0.08$. Below the critical surface is
normal state.} \label{fig.1}
\end{figure}

\begin{figure}[htbp]
\includegraphics[width=9.0cm, height=6.0cm]{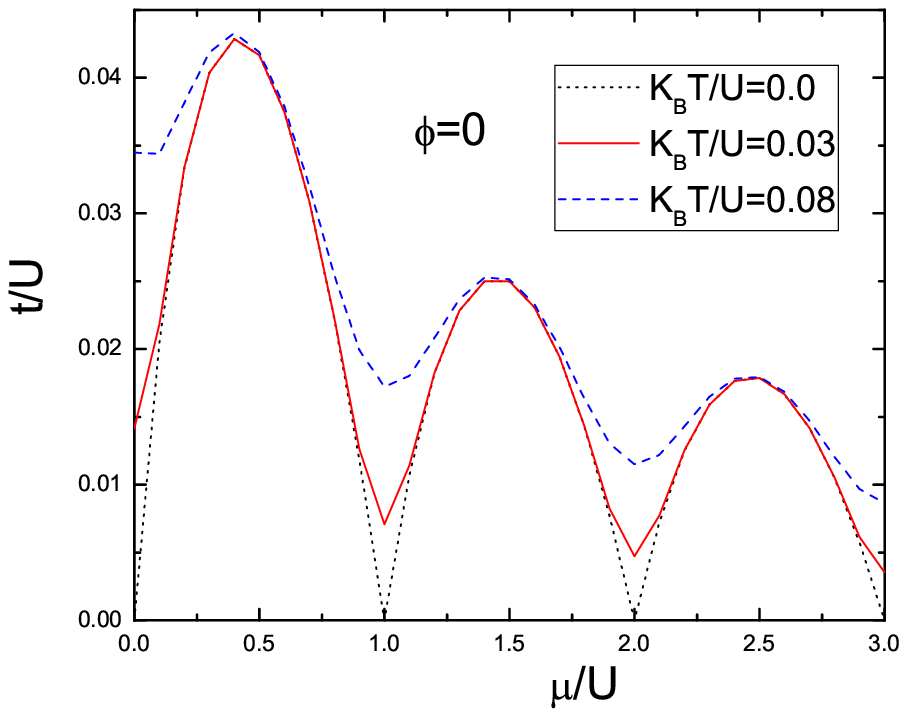}
\includegraphics[width=9.0cm, height=6.0cm]{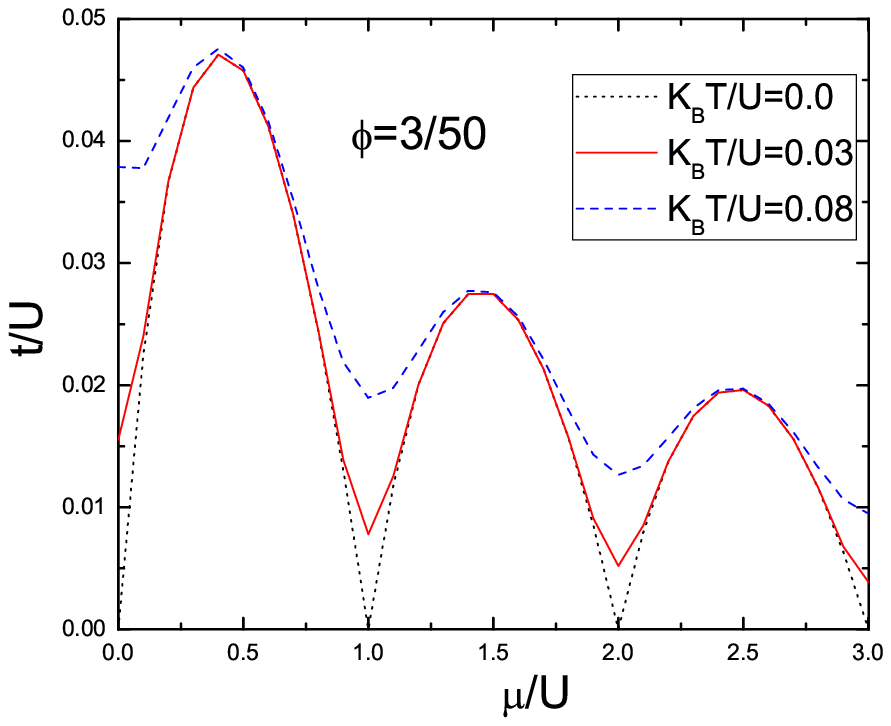}
\includegraphics[width=9.0cm, height=6.0cm]{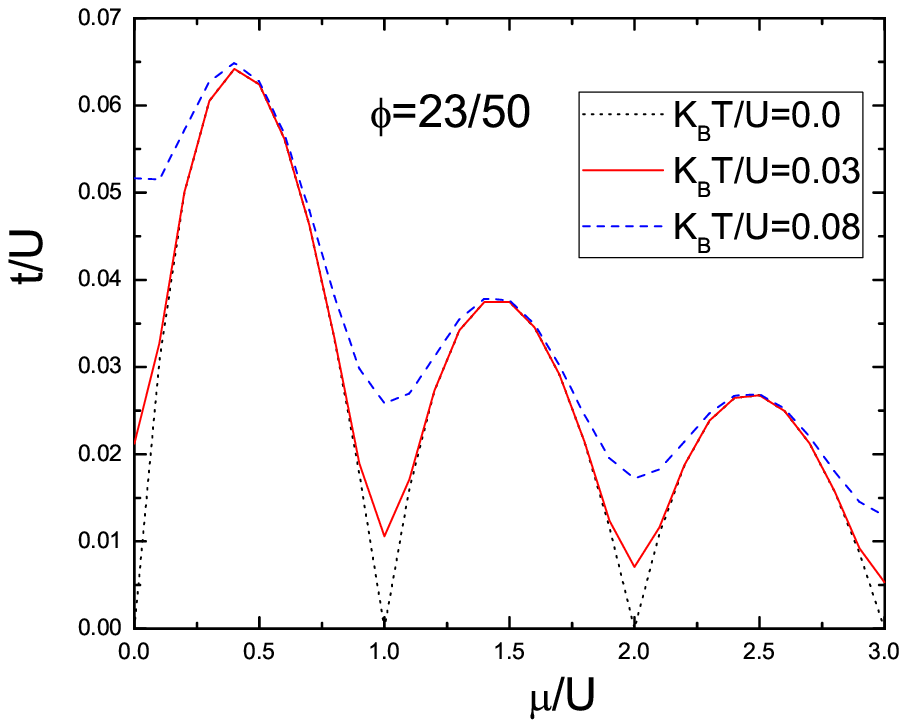}
\caption{The two-dimensional phase diagram for different temperature
and magnetic field. Below the critical line is normal state.}
\label{fig.2}
\end{figure}

In Fig.1, we plot three-dimensional phase diagram for different
temperature. When $T=0K$ the phase boundary separates the superfluid
phase and MI, which is not true for finite temperature. In order to
concentrate on the effects of temperature and magnetic field, we
also plot the two dimension phase diagram for different temperature
and magnetic field in Fig.2. Note that the phase boundary of
$\phi=0$ is obtained by letting $\epsilon_{max}(0)=4$
\cite{goldbaum, hofstadter}. Seeing from the Fig.2, we could draw
below conclusions. For fixed magnetic field the higher the
temperature, the smaller the area of the superfluid phase, which is
consistent with the fact that the high temperature destroys the
superfluidity. At the same time for fixed temperature, magnetic
field has a much more complex effect on the superfluidity in view of
oscillatory behavior of critical hopping matrix. Generally speaking
in contrast to the situation without the magnetic field, magnetic
field always nonmonotonically increases the area of normal state,
which can be illustrated from the bandwidth of Hofstadter butterfly.
On the one hand the narrower the bandwidth, the smaller the
effective hopping matrix. But the bandwidth of Hofstadter butterfly
is often less than that without magnetic field \cite{hofstadter}.
Hence the effective hopping matrix is always less than $t$. On the
other hand the on-site interaction strength is invariant.
Considering the above two factors, we naturally understand the
effect of magnetic field.

\section{Conclusion}

In conclusion we have qualitatively derived the Hamiltonian of
rotating optical lattice, analytically extended the phase diagram of
rotating Bose-Hubbard model to finite temperature and analyzed the
relation between the oscillation behavior of critical hopping matrix
and Hofstadter butterfly. In addition, we have also illustrated how
the rotation influences the phase diagram.

\section*{Acknowledgement}

The work was supported by National Natural Science Foundation of
China under Grant No. 10675108.

\begin{figure}[htbp]
\caption{The three-dimensional phase diagram for different
temperature $K_BT/U=0.0, 0.03, 0.08$. Below the critical surface is
normal state.} \label{fig.1}
\end{figure}

\begin{figure}[htbp]
\caption{The two-dimensional phase diagram for different temperature
and magnetic field. Below the critical line is normal state.}
\label{fig.2}
\end{figure}

\end{document}